\documentclass[10pt,conference]{IEEEtran}
\IEEEoverridecommandlockouts
\usepackage{amsmath,amssymb,amsfonts}
\usepackage{adjustbox}
\usepackage{graphicx}
\usepackage{textcomp}
\usepackage[colorlinks=false, hidelinks]{hyperref}
\usepackage[capitalize]{cleveref}
\usepackage[detect-all,per-mode=symbol]{siunitx}
\usepackage{booktabs}
\usepackage{multirow}
\usepackage{pgfplots}
\usepackage[protrusion=true,expansion=true]{microtype} 
\usepackage[keeplastbox]{flushend}
\usepackage[noend]{algcompatible}
\usepackage{algorithm}
\usepackage{algpseudocode}
\newcommand{\argmax}{\operatornamewithlimits{arg\,max}}

\usepackage[style=ieee, maxcitenames=2, mincitenames=1, backend=bibtex]{biblatex}
\AtEveryBibitem{
 	\clearfield{url}
 	\clearfield{doi}
\ifentrytype{inproceedings}{
 	\clearlist{address}
 	\clearlist{publisher}
 	\clearname{editor}
 	\clearlist{organization}
 	\clearfield{pages}
 	\clearlist{location}
	\clearfield{volume}
	\clearfield{series}
	\clearfield{eprint}
	\clearfield{eprinttype}
 }{}
 }

\DeclareMathAlphabet\mathbfcal{OMS}{cmsy}{b}{n}

\AtBeginBibliography{\small}

\def\BibTeX{{\rm B\kern-.05em{\sc i\kern-.025em b}\kern-.08em
    T\kern-.1667em\lower.7ex\hbox{E}\kern-.125emX}}
    
\usepgfplotslibrary{groupplots}
\usetikzlibrary{positioning}
\pgfplotsset{compat=newest}
\pgfplotsset{every axis legend/.append style={%
cells={anchor=west}}
}
\pgfplotsset{every axis/.append style={
                    label style={font=\footnotesize},
					tick label style={font=\footnotesize},
					legend style={font=\footnotesize}
                    }}

\begin{document}

\title{
    Multi-agent Reinforcement Learning with Graph Q-Networks for Antenna Tuning
}

\author{Maxime Bouton, %
    Jaeseong Jeong, Jose Outes, Adriano Mendo, Alexandros Nikou
    \thanks{The authors are with Ericsson. Email:
            {\tt \{maxime.bouton, jaeseong.jeong, jose.outes, adriano.mendo, alexandros.nikou\}@ericsson.com}}%
            
}

\maketitle

\begin{abstract}
Future generations of mobile networks are expected to contain more and more antennas with growing complexity and more parameters. 
Optimizing these parameters is necessary for ensuring the good performance of the network. 
The scale of mobile networks makes it challenging to optimize antenna parameters using manual intervention or hand-engineered strategies. 
Reinforcement learning is a promising technique to address this challenge but existing methods often use local optimizations to scale to large network deployments. 
We propose a new multi-agent reinforcement learning algorithm to optimize mobile network configurations globally. 
By using a value decomposition approach, our algorithm can be trained from a global reward function instead of relying on an ad-hoc decomposition of the network performance across the different cells. 
The algorithm uses a graph neural network architecture which generalizes to different network topologies and learns coordination behaviors.
We empirically demonstrate the performance of the algorithm on an antenna tilt tuning problem and a joint tilt and power control problem in a simulated environment. 
\end{abstract}

\begin{IEEEkeywords}
multi-agent reinforcement learning, mobile networks, 6G, graph neural networks
\end{IEEEkeywords}


\section{Introduction}

Mobile networks can be composed of thousands of base station antennas, and each of them comprises many parameters to be configured. 
The configuration of these parameters such as tilt or power usually has a great impact on the overall network performance. 
With the growing complexity of networks in 5G and beyond, the problem of dynamically configuring those parameters becomes increasingly challenging \cite{mwanje2016}.
In addition to being costly, human monitoring and intervention on the network are not scalable.
Multiple parameters can be used to affect the same performance metric, and changes in the configuration of one base station are likely to influence neighboring base stations and degrade their performance. 
In addition, the optimal choice of parameters is highly dependent on environmental factors as well as the spatial distribution of the traffic and the mobility of the connected user equipment. 
There is a need to design algorithms that can automatically learn tuning strategies to optimize the network performance by adapting to changes in the environment while considering all the possible interactions across neighboring base stations and across parameters.

In contrast to hand-engineered or manual tuning strategies, reinforcement learning (RL) provides a flexible framework to learn a control strategy from data. 
Previous works have demonstrated its application to a variety of radio access network optimization use cases including remote electrical tilt control, optimization of handovers in 5G or power control, including industrial solutions~\cite{vannella2022,yajnanarayana2020, Ghadimi2017}. 
Although they are promising, existing RL methods often fail to capture the needed coordination across neighboring base stations. 
For scalability reasons, they often resort to limiting assumptions and consider the control of one entity independently of the others, or rely on specific feature engineering to incorporate neighbor information as input to the reinforcement learning agents. 
In addition, they focus on optimizing local Key Performance Indicators (KPIs) involving one base station and its closest neighbors instead of considering the network as a whole in the reward formulation. 
In this work, we derive a multi-agent reinforcement learning algorithm capable to optimize network performance globally and control many base stations in a coordinated manner without relying on local reward approximation and feature engineering. 

Our approach extends state-of-the-art cooperative multi-agent reinforcement learning algorithms by proposing a novel off-policy algorithm for cooperative learning when a graph structure is available. 
Inspired by value decomposition methods~\cite{sunehag2018,rashid2018}, we propose a new Q-network architecture, graph Q-network (GQN), that is particularly well suited to address mobile network optimization problems. 
GQN uses graph neural networks to generalize to different network topologies and share knowledge across neighboring base stations and different parameters to control.
Our GQN training algorithm uses a reward signal characterizing the performance of the network in a global area as opposed to using ad-hoc credit assignment techniques, and thereby, the agents learn to coordinate to improve the global network performance. 
We demonstrate the performance of the proposed algorithm on two mobile network optimization problems: tilt control and joint tilt and power control. 
In the first problem, we show that GQN outperforms all the baselines and can generalize to different network topologies. 
In the second problem, we show the ability to control different antenna parameters in a scalable way by considering each parameter as an agent.
A technical report with additional details and experiments in appendix can be found at \url{preprint url}.


\section{Related Work}

Existing works applying reinforcement learning for antenna tuning consider only local information of a cell and its closest neighbors~\cite{balevi2019online, farooq2019ai}. 
Several methods have been proposed to do so. 
The first one consists in hand-engineering an input feature and reward function to accommodate for the effect one cell can have on its neighbors.
The KPIs of the neighboring cells are aggregated and the RL agent tries to optimize a combination of its own KPIs and the neighbor's KPIs~\cite{vannella2021}. 
Other techniques have proposed to use graph neural networks to process neighbor information. 
However, they also require an ad-hoc engineering of the reward and can only control one base station at a time~\cite{jin2022}. 
Other algorithms attempting to address the global network optimization problem have been proposed in previous works using coordination graphs~\cite{bouton2021}.
This solution also required a heuristic to handle credit assignment between base stations by splitting individual rewards across neighbors. The inference cost of the message passing algorithm is larger than in our proposed method as it requires storing a neural network for every connected base stations in the graph.
Our proposed algorithm can train a model controlling multiple antennas from a single global reward signal. 
In addition, previous works focus on controlling one antenna parameter, in this work we demonstrate the ability to control multiple parameters simultaneously (tilt and power). 

Multi-agent RL algorithms have been proposed for problems where multiple agents cooperate for a common goal. The closest algorithms to our methods are value decomposition networks (VDN) and QMIX~\cite{sunehag2018, rashid2018}. 
They rely on factorizing the problem to only learn one value function per agent contrary to using individual reward signals. They train these individual value functions using one global reward such that the sum of individual values (or a weighted sum in the case of QMIX) matches the global reward. We take inspiration from these factorization methods but add a graph neural network component. It can exploit the topology of the telecommunications network to learn a more efficient decomposition of the joint action value function. 
Algorithms in the literature have not been applied to network optimization problems and have rarely been scaled to more than dozens of agents. 
Our algorithm scales to hundred of agents and can cope with a varying number of agents both at training and deployment time.

\section{Background}

\subsection{Multi-agent Reinforcement Learning}\label{sec:marl}

The problem of network optimization can be formulated as a multi-agent cooperative reinforcement learning problem where each network entity is an agent. 
The problem is modeled as a multi-agent Markov decision process~\cite{dmu}.
Formally, it is described by the tuple $(\mathbfcal{S}, \mathbfcal{A},T,R, \gamma)$, where $\mathbfcal{S}$ is a joint state space, $\mathbfcal{A}$ a joint action space, $T$ an unknown transition model, {$R: \mathbfcal{S}\times\mathbfcal{A}\rightarrow \mathbb{R}$}  a global reward function, and $\gamma\in[0,1)$ a discount factor. 
A joint state $\mathbf{s}\in\mathbfcal{S}$ is equal to $(s_1,\ldots,s_n)$ where $s_i$ is the state of agent $i$ and $n$ is the number of agents. 
Similarly, a joint action $\mathbf{a}\in\mathbfcal{A}$ is equal to $(a_1,\ldots,a_n)$ where $a_i$ is the action of agent $i$. The agents do not need to be homogeneous and could have different action spaces. 
Our goal is to find a joint policy $\pi: \mathbfcal{S}\rightarrow\mathbfcal{A}$ that maximizes the discounted accumulated global reward over time. 
A standard approach in RL is to represent $\pi$ by a value function $Q: \mathbfcal{S}\times\mathbfcal{A}\rightarrow \mathbb{R}$  such that $\pi(\mathbf{s}) = \argmax_\mathbf{a} Q(\mathbf{s}, \mathbf{a})$ . 
In single-agent cases, the Q-learning algorithm can be used to learn $Q$. For problems with continuous, or large state spaces, $Q$ can be represented by a neural network and learned using deep Q-learning~\cite{mnih2015}. It minimizes the Bellman error: 
\begin{equation}
    \min_\theta E_{(\mathbf{s},\mathbf{a},r,\mathbf{s'})}[(Q(\mathbf{s}, \mathbf{a}; \theta) - (r + \gamma \max_{\mathbf{a}'}Q(\mathbf{s'}, \mathbf{a}'; \theta))^2]
    \label{eq:Bellman}
\end{equation}
where $(\mathbf{s}, \mathbf{a}, r, \mathbf{s}')$ is a multi-agent experience sample collected by interacting with the environment and $\theta$ are the weights parameterizing the value function. 
The reward is a global reward for the whole system. 

Representing and estimating $Q$ in multi-agent systems is hard because of the dimensionality of the joint state and action spaces increasing with the number of agents. 
In addition, solving the maximization problem to find $\pi(s)$ is also challenging due to the large size of the joint action space. 
A common approach is to approximate the value function by factorizing it into individual value functions as follows~\cite{sunehag2018}: $Q(\mathbf{s}, \mathbf{a}) \approx \sum_{i=1}^n Q_i(s_i, a_i)$
where $Q_i$ is the individual value function associated to agent $i$. 

\subsection{Graph Neural Networks}

Graph neural networks (GNNs) are a family of neural network architectures designed to process graph structured data. 
Generally, a GNN is a differentiable parametric function that takes as input a graph with node and edge attributes. 
Consider a graph $\mathcal{G}$ with vertices $\mathcal{V}$ and edges $\mathcal{E}$ where each node $i\in\mathcal{V}$ is associated to an attribute vector $x_i\in\mathbb{R}^d$. 
A simple graph neural network layer processing the input graph $\mathcal{G}$ can be described by the following equation~\cite{morris2019}: $h_i = \sigma \left( W_1x_i + W_2\sum_{j\in\mathcal{N}(i)} x_j \right) \label{eq:gnn}$
where $h_i$ is the latent features of node $i$, $\sigma$ is an activation function, $W_1$ and $W_2$ are learnable weight matrices, and $\mathcal{N}(i)$ represents the set of neighbors of vertex $i$ in the graph. 
Other types of GNNs have been considered with different ways of aggregating neighbor features and considering edge features such as using convolution or attention like operators~\cite{kipf2017,velickovic2018,morris2019}.

\section{Proposed Approach}

In this section we describe how to model network optimization problems using cooperative multi-agent reinforcement learning. 
We then introduce a new algorithm that can efficiently exploit the inherent graph structure of the network to automatically coordinate different agents even in a scenario where they control different types of base station parameters.

\subsection{System Model}

Mobile networks are composed of several base stations equipped with antennas responsible for serving users in certain areas. 
An area of coverage and its associated antenna in a radio access network is referred to as a cell. 
In 5G and 6G networks, the number of cells in a network is expected to increase drastically by combining macro cells responsible for covering a large area and small cells providing increased capacity in targeted zones.
Each of the antennas broadcasting a signal to the cell can be tuned to maximize the signal quality which results in improved quality of experience for the users of that cell. 
In this work we focus on two different antenna tuning scenarios: tuning the remote electrical tilt, and jointly tuning the tilt and the downlink transmission power. 
Our goal is to demonstrate that our proposed algorithm can handle environments with heterogeneous parameters to be controlled.

Several metrics can be considered to maximize the user quality of experience such as throughput, coverage or signal quality. 
In this work, we focus on maximizing the downlink signal quality of all the users in the network as measured by the reference signal to interference and noise ratio (SINR) in \SI{}{\decibel}. 
The goal of our algorithm is to find a policy mapping the global state of the network to a joint cell configuration in order to maximize the SINR of the whole network. 

The SINR, $\rho_u$ of a user $u$ in a cell $c$ depends on the tilt angle $\alpha$ and the downlink power $\phi$ as follows {$\rho_{u}(\mathbf{\phi}, \mathbf{\alpha}) = \frac{R_{c_u}(\phi_c, \alpha_c)}{\sum_{i=1,i\neq c}^{N_{\text{cells}}} R_{i,u}(\phi_i, \alpha_i) + \mu}$} where $R_{c,u}$ is the reference signal received power (RSRP). The relation between RSRP and antenna parameters is derived through antenna models defined in standards~\cite{3gpp.36.814}. More detailed on the RSRP calculation can be found in the appendix C of our technical report.

We consider the global SINR of the network as the average of the SINR of all the users in the network: $\text{SINR}_G(\pmb{\alpha}, \pmb{\phi}, \mathbf{s}) = 1/N_{\text{users}} \cdot \sum_u \rho_{u}$. Considering a geometric mean would have also been possible to model fairness among the users.
The global SINR is a function of the joint configuration of all the cells in the network, along with some other exogenous factor such as the environment and the traffic distribution, all abstracted in the variable $\mathbf{s}$.
In constrast to the global SINR, we define a local SINR for a given cell $c$: $\text{SINR}_{L,c} = 1/N_c \cdot \sum_{u\in c} \rho_{u}$  which is the average SINR of all the users connected to that cell, where $N_c$ is the number of users connected to cell $c$.

\subsection{Global Network Optimization as a Multi-agent MDP}\label{sec:model}

\begin{figure}
    \centering
    \includegraphics[width=0.8\columnwidth]{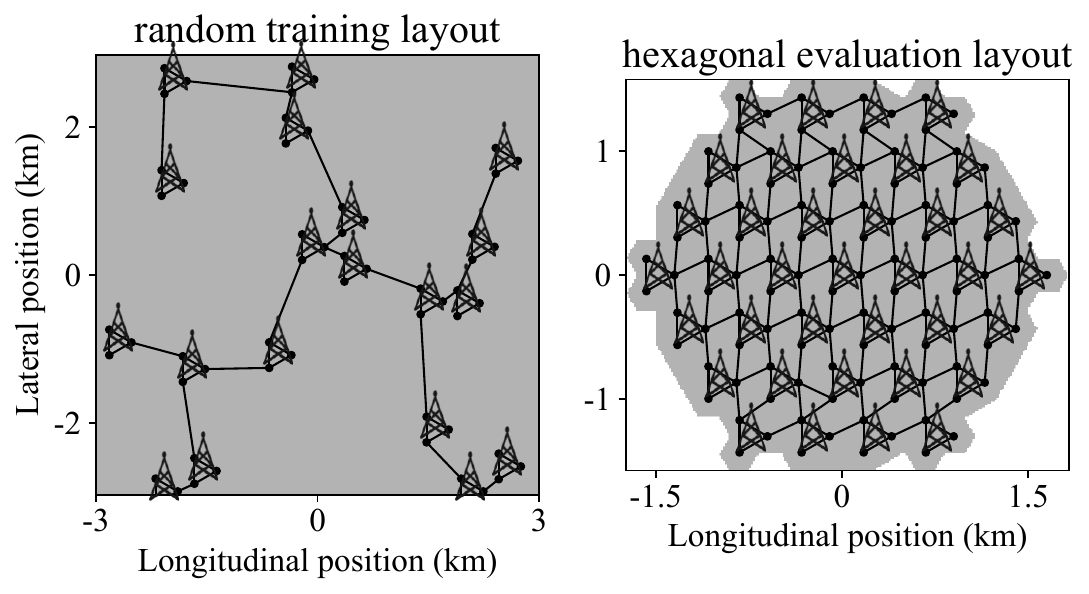}
    \caption{Macro base station network deployments used in our experiments with the graph connecting the different cells (three per base station). The topology of the graph can change depending on the intersite distance. Evaluation layouts use denser deployments with more agents. The position of the node is offset from their real position for better visualization.}
    \label{fig:deployment}
\end{figure}

Maximizing the global network performance can be formulated as a multi-agent Markov decision process where each cell in the network is an agent. 
We assume that a graph modelling the relation between the different cells is available. 
Each node is a cell, and a vertex exists between two cells if they can influence each other. 
Automatic neighbor relations methods defined in the standards can be used to identify these edges~\cite{3gpp.36.814}. 
In our simulation, we use a criterion based on the geographical location and the azimuth angle of each cell along with the interference ratio between the cells. 
Namely, if cells are close to each other and have high mutual interference they are connected in the graph. 
An example of such a graph is illustrated in \cref{fig:deployment}.

Similarly to previous work~\cite{jin2022}, each agent $i$ is assumed to observe the following quantities: the position of the antenna $(x_i, y_i)$, the direction of the antenna $(x^\Delta_i, y^\Delta_i)$, the 10th, 50th, and 90th percentile of the SINR of the connected users, the antenna tilt angle $\alpha_i$, and the antenna maximum downlink power $\phi_i$.
The joint state space $\mathbfcal{S}$ is represented by the Cartesian product of the individual state of each cell. 

We now consider two different problems: a tilt control problem and a joint tilt and power control problem. 
Each problem has a different action space and a different reward function.
The true objective of these antenna tuning problems is to maximize the global SINR of the network to improve the quality of service and minimize the transmitted power to reduce the energy consumption (in the case where power is controllable). 
Existing multi-agent RL algorithms often must rely on ad-hoc decompositions of the reward signal in order to scale to a large number of agents.
In order to compare with these baselines, we define both a global reward function that can be used by our method and a local reward function that will be used by the proposed baseline algorithm in our experimental section.
The global SINR is defined as the arithmetic mean of the SINR of all the users in the network. However, other definitions such as geometric mean could be used in order to promote fairness. 
Our method could be applied in a straightforward way to these other definitions.

\textbf{Tilt control:}
The tilt action space is defined by a set of remote electrical tilt changes of $\{\SI{-1}{\degree}, \SI{0}{\degree}, +\SI{1}{\degree}\}$. The tilt is always bounded within the range $[\SI{0}{\degree},\SI{15}{\degree}]$. In this scenario, the maximum transmission power of each cell is fixed to $\SI{40}{\watt}$.
The local and global reward functions are defined as follows:
\begin{itemize}
    \item Global SINR reward: $R(\mathbf{s}, \mathbf{a}, \mathbf{s'}) = \text{SINR}_G(\pmb{\alpha}, \pmb{\phi}, \mathbf{s})$. 
    \item Local SINR reward for cell $i$: $R_i(\mathbf{s}, \mathbf{a}, \mathbf{s'}) =  \text{SINR}_{L,i} + 1/N_i \cdot \sum_{j\in\mathcal{N}(i)}\text{SINR}_{L, j}$. This local formula considers the performance of a single cell and the average performance of its neighbors. We consider all the neighbors in the graph and each of them are equally weighted. 
    A disadvantage of this approach is that we need to choose how to weigh the neighbor manually, while our proposed approach will learn automatically the reward decomposition.
    Similarly to the formula above, we only consider tilt when using this formula. 
\end{itemize}

\textbf{Joint tilt power control:}
The joint action space is defined by the Cartesian product of the tilt action space described above, and a set of maximum downlink power changes $\{\SI{-5}{\watt}, \SI{0}{\watt}, +\SI{5}{\watt}\}$ resulting in a total of \num{9} actions per cell.
The power is bounded to the range $[\SI{10}{\watt}, \SI{60}{\watt}]$.
The reward functions are defined by:
\begin{itemize}
    \item Global SINR and power reward $R(\mathbf{s}, \mathbf{a}, \mathbf{s'}) = (1 - w) \text{SINR}_G(\pmb{\alpha}, \pmb{\phi}, \mathbf{s}) -  w \cdot 1/N_{\text{cells}} \cdot \sum_c\phi_c$. This reward definition adds a penalty based on the average transmitted power of the cells. The penalty is weighted by a hyperparameter $w$. Setting $w$ to \num{0} would yield to maximizing the power (since it leads to maximum SINR) and optimizing tilt, setting $w$ to \num{1} would lead to completely minimizing the power and disregarding any effect of the tilt. In practice, the choice of $w$ should be driven by business intents from the network operators. 
    \item Local SINR and power reward $R_i(\mathbf{s}, \mathbf{a}, \mathbf{s'}) =  (1 - w)\text{SINR}_{L, i} - w\phi_i + 1/N_i \cdot \sum_{j\in\mathcal{N}(i)}[(1-w)\text{SINR}_{L, j} - w\phi_j]$
\end{itemize}
Given a joint network state $\mathbf{s}$, we are interested in a policy giving a joint configuration of tilt and power to maximize the expected accumulated reward using the formalism of multi-agent reinforcement learning described in \cref{sec:marl}. 
The number of actions per cell grows exponentially with the number of parameters considered. 
Since we consider only tilt and power, the size of the action space is still reasonably small and we can learn a single joint controller for tilt and power. 
In our experiments, we are also going to consider the possibility of making two separate agents for each cell, one controlling the tilt and one controlling the power.

Maximizing the global reward is generally challenging due to the complex interactions between the cells in the networks. 
Instead one can rely on value decomposition techniques such as independent Q-learning~\cite{tan93}.
Each cell would be learning using their local observation and the proposed local reward definition considering only direct neighbors. 
The local reward ignores the fact that some neighbors might be more important than other and that indirect interactions with neighbors further away in the graph could affect performance. 
To consider the global effect of all the cell in the network, we propose the graph Q-network algorithm.

\subsection{Graph Q-Network}

We describe a multi-agent reinforcement learning algorithm to control multiple cells in the network optimizing a single global objective. 
The algorithm relies on graph neural networks to process information from all the agents. 
It then uses a factorization technique to learn individual value functions for each cell from a single global reward signal, rather than relying on hand-engineered reward decomposition across the cells.
At deployment time, the trained model can be evaluated with an arbitrary number of cells allowing to train at small scale but deploy at large scale. 
These properties are mostly enabled by an original representation of the joint state action value function which we name graph Q-network (GQN). 

To train such a model using a single global reward signal for all the agents, we rely on a factorization technique. 
Factorization generally consists of decomposing a very large function into a combination of smaller components. 
In this problem, we wish to learn an additive decomposition of the joint state action value function such that: $Q(\mathbf{s}, \mathbf{a}) = \sum_i Q_i(h_i, a_i)$ where $(h_1, \ldots, h_n) = \text{GQN}(\mathbf{s},A)$.
The GQN function represents our proposed model and $A$ denotes the adjacency matrix of the network graph. 
Intuitively, the GQN is going to learn a hidden state representation for each agent, such that the global value function is linearly factorizable into individual value functions in that hidden space. To simplify the notation, we note each output node of the GQN as $Q_i(\mathbf{s}, A, a_i)$, which represents the value of taking action $a_i$ for node $i$. 

\begin{figure}
    \centering
    \includegraphics[width=\columnwidth]{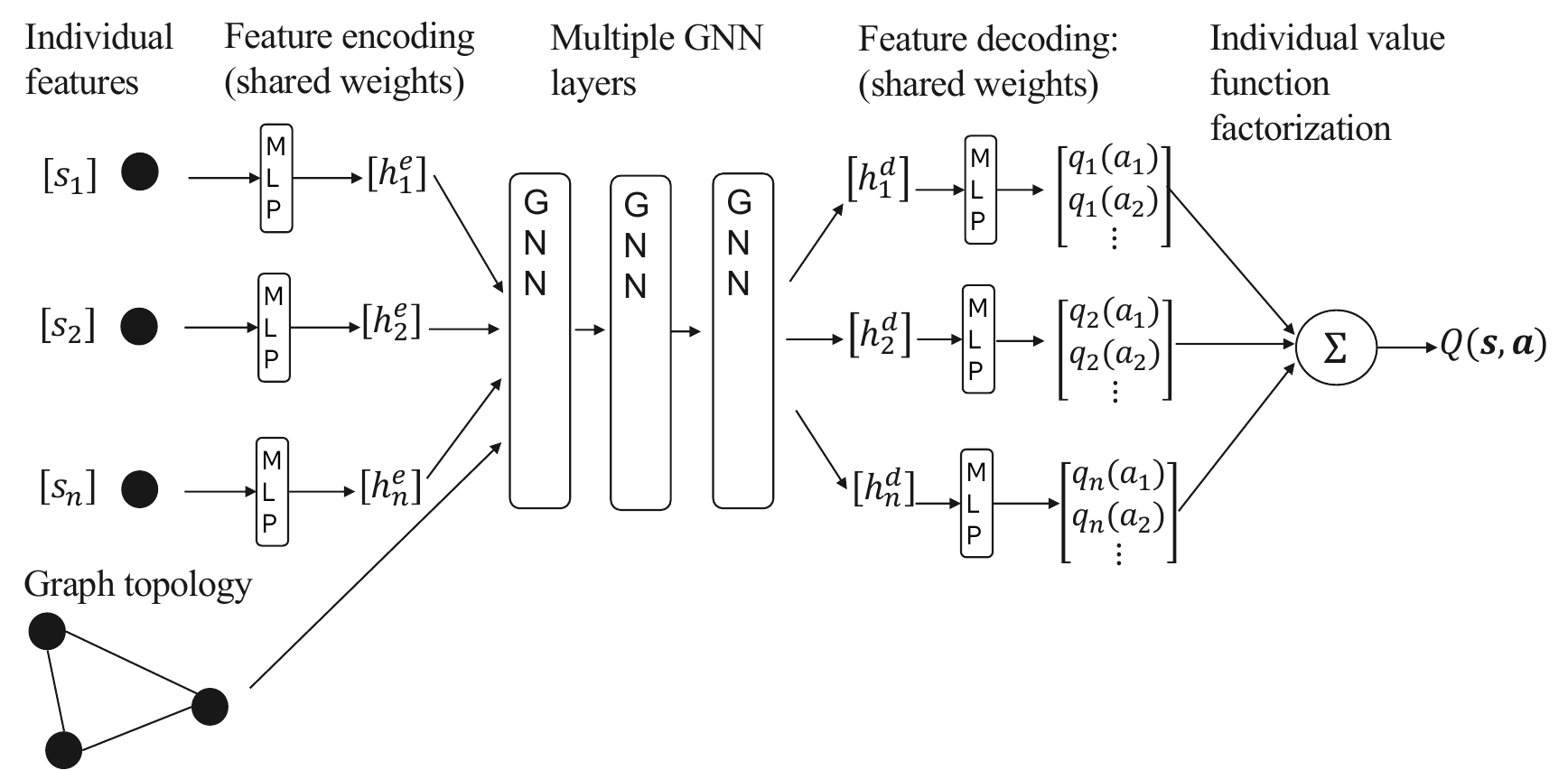}
    \caption{Graph Q-network architecture}
    \label{fig:gqn-architecture}
\end{figure}

The GQN architecture is illustrated in \cref{fig:gqn-architecture}. 
It is parameterized by weights and is end to end differentiable.
The model takes as input a feature vector for each agent to control along with a graph representation of the mobile network. 
The construction of such graph is described in the previous section and has been also demonstrated in previous work~\cite{bouton2021,jin2022}. 
Each agent is represented by a node in the graph.
The feature vector $s_i$ corresponds to the local state of a cell as described in the previous section. 
It is processed by an encoding layer which consists of multilayer perceptrons (MLPs) applied individually to each node feature vector. 
To improve sample efficiency during training and the generalization of the model to an arbitrary number of nodes, the MLPs have shared weights. 

The encoded features, $h_i^e$, are then processed by several GNN layers. 
We experimented with the GNN architectures from \citeauthor{morris2019}~\cite{morris2019} and the graph attention neural networks from \citeauthor{velickovic2018}\cite{velickovic2018}. 
The output of the GNN layers consists of another set of hidden features, $h_i^d$, one for each node, passed through a decoding layer.
The model is trained such that the output node embeddings represent the individual value function of each agent which we note as: $Q_i(h^d_i, a_i)$. 

Training a GQN model follows an off-policy training procedure similar to DQN and can benefit from existing innovations like prioritized experience replay, target networks and double Q learning~\cite{hessel2018}. 
The convergence guarantees of GQN are similar to the one provided by DQN. The training procedure of GQN is described in our technical report.

The GQN architecture has the sufficient property (additivity) to satisfy the individual global maximum principle~\cite{son2019} which states that the jointly optimal actions are given by taking the individual maximum of each individual value function at the output of the GQN model:
\begin{equation}
\mathbf{a^*} = \argmax Q(\mathbf{s}, \mathbf{a}) = \begin{bmatrix}
           \argmax Q_1(h^d_1, a_1) \\
           \argmax Q_2(h^d_2, a_2) \\
           \vdots \\
           \argmax Q_n(h^d_n, a_n)
         \end{bmatrix}
\label{eq:igm}
\end{equation}
Contrary to previous work~\cite{jin2022}, the model outputs state-action value functions for all the agents at the same time. 

\section{Experiments}\label{sec:experiments}

We empirically demonstrate the performance of our proposed algorithm in simulated environments. 
Using the scenarios in \cref{sec:model}, we demonstrate three different experiments illustrating respectively: the training performance for tilt environment, the generalization performance for tilt control, the training performance for the joint control problem with an additional agent decomposition scheme.

\subsection{Experiment Setup}

Our experiments rely on a proprietary system level simulator which implements a ray tracing propagation model for computing the path gains at various user locations~\cite{asplund2018}. Details about the simulation parameters and training parameters can be found in our technical report in appendix\footnote{Experimental details can be found in appendix A here: \url{technical report}}.

 We compare the following algorithms: 
\begin{itemize}
    \item DQN: The DQN algorithm with a target network, double Q-learning, prioritized replay, and distributed training~\cite{hessel2018}. 
    \item Neighbor DQN (N-DQN): it is a simple extension to DQN where the observation is augmented with the observation of the neighboring cells, proposed in previous work~\cite{jin2022}.
    \item GAQ~\cite{jin2022}: the graph attention Q-network algorithm processes neighbor features using a graph attention layer. Contrary to our proposed approach, GAQ is trained using a local reward and is not able to perform joint control. 
    \item GQN: our proposed method. We write GQN(GAT) when we used a graph attention layer~\cite{velickovic2018}, otherwise we used a graph convolutional layer from \citeauthor{morris2019}~\cite{morris2019}.
    \item Heuristic (H): This method is a rule-based method that sets the tilt angle such that the beam points to the middle of the cell. It observes the height of the antenna and the distance to its closest neighboring cell and calculates a desired tilt angle. 
\end{itemize}

We did not add QMIX as part of the baselines as it is already shown to be outperformed by GAQ in previous work~\cite{jin2022}. 
On the joint tilt and power control problem, QMIX was at most as good as DQN for some seeds, but the training was too unstable to report meaningful results in this paper.
In the figures, the solid lines represent the mean over \num{3} random seeds and the shaded area is the \num{95}th percentile confidence interval.
    The line for the heuristic represents the average performance and confidence interval evaluated on 300 episodes.

\subsection{Training performance}

\begin{figure}
    \centering
    \input{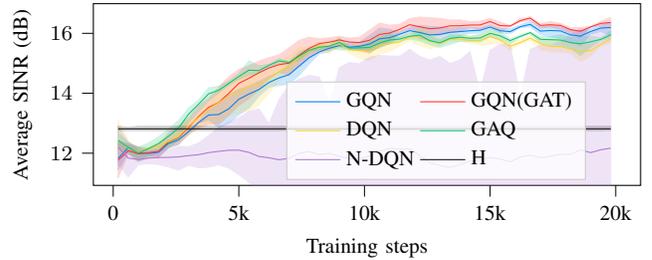}
    \caption{Evolution of the average SINR per episode during the training of the agents for tilt control. We compare our approach (GQN, GQN(GAT)) to a standard DQN approach, N-DQN, and GAQ, a state-of-the-art algorithm developed for antenna tilt control~\cite{jin2022}.}
    \label{fig:tilt-training}
\end{figure}

\Cref{fig:tilt-training} illustrates the performance of GQN, GQN(GAT), DQN, GAQ and N-DQN on a tilt tuning scenario. We report the average SINR of the whole network (averaged over all the user equipments) which corresponds to the global SINR optimization target. 
The results are shown when training in a hexagonal deployment. 
A training step corresponds to one interaction with the environment. 
Similar conclusions can be drawn when training in the scenario with random deployments.
We can see that all the methods except from N-DQN have similar convergence properties. In these experiments, all the methods follow DQN-like procedure and the convergence was mostly affected by the choice of the exploration schedule which is the same for all the baseline for a fair comparison.
For the heuristic, we plot the average performance, as it is not a learning based algorithm and the performance is constant. We can see that all the RL algorithms outperform the baseline except for N-DQN.

While almost all the methods converge to an improved average SINR compared to the initial performance, GQN provides the best solution. 
We notice a difference of about \SI{1}{\decibel} in the average performance. 
The convergence rate is roughly similar for all the methods with a slight disadvantage for GQN with the graph convolutional layer. 
The neighbor DQN approach performed poorly in general. We notice that the poor performance is accompanied by a high standard deviation. This is explained when looking to each of the three random seeds, only one performed at a level comparable to GAQ and DQN while the other seeds did not converge. We blame this instability on the complexity of the observation space and the difficulty to process this information with a simple fully connected neural network. 
The complexity of adding the graph neural network layers does not seem to affect the convergence for both GQN and GAQ. 
This result confirms the initial intuition that being able to train on the global objective rather than manually decomposing the global reward into individual cell reward (for DQN and GAQ) leads to better global performance.

\subsection{Generalization to New Deployments}

In this experiment, we save the trained models from \cref{fig:tilt-training} and evaluate them on network deployments unseen at training time. 
In addition, we evaluate another batch of models that were trained on random deployments rather than hexagonal deployments. 
We have two possible training configurations, hexagonal or random, and two evaluation configurations, hexagonal or random. Examples of a random training configuration and a hexagonal evaluation configuration are given in \cref{fig:deployment}. 
We evaluate the models with up to \num{37} macro base stations which makes a total of \num{111} agents against \num{57} at training time. 

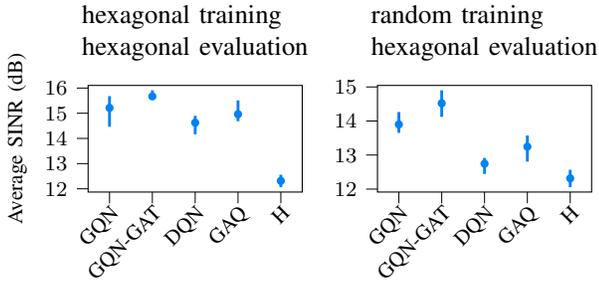
\begin{figure}
    \centering
\begin{tikzpicture}

\definecolor{darkgray176}{RGB}{176,176,176}
\definecolor{dodgerblue0130240}{RGB}{0,130,240}

\begin{groupplot}[group style={group size=2 by 1,vertical sep=2.2cm}, width=0.5\columnwidth, height=3cm]
\nextgroupplot[
tick align=outside,
tick pos=left,
title style={align=left},
title={hexagonal training \\ hexagonal evaluation},
unbounded coords=jump,
x grid style={darkgray176},
xmin=-0.5, xmax=4.5,
xtick style={color=black},
xtick={0,1,2,3,4,5},
xticklabels={GQN,GQN-GAT,DQN,GAQ,H},
x tick label style={rotate=45, anchor=north east, inner sep=0mm},
y grid style={darkgray176},
ylabel={Average SINR (dB)},
ymin=11.8712364860283, ymax=16.1040258501573,
ytick style={color=black}
]

table{%
x  y  draw  fill
0 15.2123838276273 0,130,240 0,130,240
1 15.6683761327389 0,130,240 0,130,240
2 14.6257836637029 0,130,240 0,130,240
3 14.9621204331694 0,130,240 0,130,240
4 12.314558025762 0,130,240 0,130,240
};
\addplot [only marks, mark size=1pt,mark=*, line width=1.08pt, dodgerblue0130240]
table {%
0 15.2123838276273
1 15.6683761327389
2 14.6257836637029
3 14.9621204331694
4 12.314558025762
};
\addplot [line width=1.08pt, dodgerblue0130240]
table {%
0 14.4633371385209
0 15.6787333835573
};
\addplot [line width=1.08pt, dodgerblue0130240]
table {%
1 15.5383183186699
1 15.911626333606
};
\addplot [line width=1.08pt, dodgerblue0130240]
table {%
2 14.1652532386594
2 14.8982876198943
};
\addplot [line width=1.08pt, dodgerblue0130240]
table {%
3 14.6838843464045
3 15.5116990049138
};
\addplot [line width=1.08pt, dodgerblue0130240]
table {%
4 12.0636360025796
4 12.5584725724441
};

\nextgroupplot[
tick align=outside,
tick pos=left,
title style={align=left},
title={random training \\ hexagonal evaluation},
unbounded coords=jump,
x grid style={darkgray176},
xmin=-0.5, xmax=4.5,
xtick style={color=black},
xtick={0,1,2,3,4,5},
xticklabels={GQN,GQN-GAT,DQN,GAQ,H},
x tick label style={rotate=45, anchor=north east, inner sep=0mm},
y grid style={darkgray176},
ymin=11.9134688578056, ymax=15.044383670695,
ytick style={color=black}
]
table{%
x  y  draw  fill
0 13.8998679475545 0,130,240 0,130,240
1 14.5202266488269 0,130,240 0,130,240
2 12.7471026397811 0,130,240 0,130,240
3 13.246662925662 0,130,240 0,130,240
4 12.314558025762 0,130,240 0,130,240
};
\addplot [only marks, mark size=1pt,mark=*, line width=1.08pt, dodgerblue0130240]
table {%
0 13.8998679475545
1 14.5202266488269
2 12.7471026397811
3 13.246662925662
4 12.314558025762
};
\addplot [line width=1.08pt, dodgerblue0130240]
table {%
0 13.6514784774299
0 14.2640713587606
};
\addplot [line width=1.08pt, dodgerblue0130240]
table {%
1 14.1219356461126
1 14.9020693610182
};
\addplot [line width=1.08pt, dodgerblue0130240]
table {%
2 12.4445118843485
2 12.9143744870467
};
\addplot [line width=1.08pt, dodgerblue0130240]
table {%
3 12.8077595972763
3 13.5733008337825
};
\addplot [line width=1.08pt, dodgerblue0130240]
table {%
4 12.0557831674824
4 12.5667927648893
};

\end{groupplot}

\end{tikzpicture}
    \caption{Generalization performance of the GQN algorithm and the baselines on deployments unseen at training time. The points represent the average performance and the error bars represent the 95th percentile confidence interval (sometimes smaller than the marker).}
    \label{fig:generalization}
\end{figure}

\Cref{fig:generalization} illustrates the performance of the trained model on the hexagonal evaluation scenarios when training with random or hexagonal deployments. 
The performance is average over \num{50} episodes and across three random seeds. We can see that the GQN methods provide overall better performance than the other methods, followed closely by the GAQ algorithm. 
Using a graph attention layer in GQN provides an additional gain in performance. 
An interesting outcome is that the GQN(GAT) model trained on random deployment, has a comparable performance on the hexagonal evaluation scenario (right plot) than the DQN and GAQ models that were trained specifically on hexagonal topologies (left plot). 
When the difference between training deployment and evaluation deployment is large (from random to hexagonal), the heuristic has a very close performance to some of the reinforcement learning algorithms, indicating that generalization to those scenarios is a difficult task. 
We omitted N-DQN from the figure for readability, its performance is around \SI{9}{\decibel} in all cases. 
A poor performance is expected for this baseline as the neighbor features are stacked in a single vector. This input does not take into account the graph structure and is dependent on the order of the neighbors are stacked in. 

\subsection{Joint Control Problem}

In this experiment, we consider the possibility to control heterogeneous agents.
In the simplest case, the joint tilt and power control can be addressed by designing one agent per cell with an action space of size \num{9}. 
This is the approach we will refer to as ''joint``. 
Such approaches might not be scalable if the number of parameters to control grows. 
Instead, one could consider two agents per cell: one controlling the tilt, and one controlling the power, each with an action space of size~\num{3}. 
This is the approach we use in the GQN algorithm in this experiment. 
Instead of controlling \num{57} agents, the GQN will control \num{114} agents. 
We simply modify the graph topology such that each cell consists of two nodes connected to each other, but with the same neighbors as in the joint control case.

\begin{figure}
    \centering
    \input{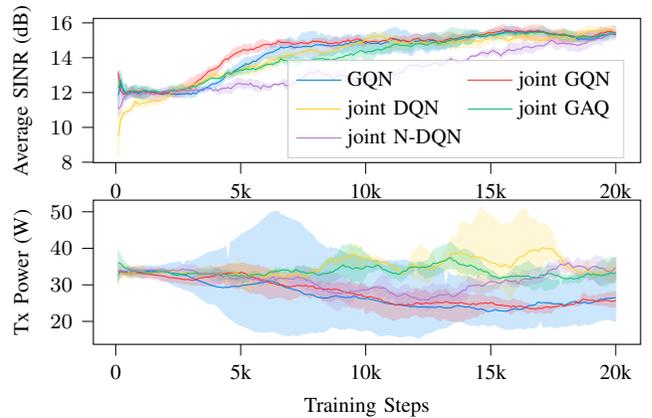}
    \caption{Training performance of GQN with separate agents for tilt and power, joint GQN and the baselines in the joint control environment. For a similar average SINR, GQN is able to yield a greater power reduction.}
    \label{fig:power-tilt-training}
\end{figure}

Results are presented in \cref{fig:power-tilt-training}. 
By looking at the two objectives, SINR and average power, we can see that the final SINR is similar for all the methods while the final power value shows a significant difference. 
For a similar average SINR, the GQN algorithm is able to reduce the power by more than \SI{35}{\percent}.
In addition, the GQN methods tend to converge faster on this problem. 
Our hypothesis is that the increase of dimensionality in the action space impacts more the methods relying only on fully connected neural networks.

Another observation is that the GQN approach yields the same average performance as joint GQN. 
This result suggests that GQN might be a good algorithm to decompose the antenna tuning problem into multiple agents per parameters. 
We expect that the standard deviation could be further reduced by more intense hyperparameter tuning. 
 
\section{Conclusions}

We presented a novel multi-agent reinforcement learning solution to tune many antennas in mobile networks. 
The proposed algorithm is an off-policy algorithm that, given a reward associated to the global network performance, learns to assign credits to the different antennas. 
Our method is able to learn the credit assignment and coordination behavior thanks to a graph Q-network, a graph neural network representation of the joint state action value function. 
Our experiments evaluate the learning algorithm in a tilt control scenario and a joint power and tilt control scenario. 
The results show that the trained model is able to learn control policies leading to a better global average SINR and more power savings. 
In addition, the trained GQN is able to generalize to denser network topologies unseen at training time, with almost double the number of agents. 
In the joint control scenario, we illustrated a way to separate the control of two parameters as different agents connected in a graph. 
By splitting the agents per parameter it allowed to maintain a smaller action space per agent while leading to similar performance as a joint controller.

Future works involve evaluating the algorithm on a broader range of problems and learning the graph structure automatically. We also consider investigating the control of advanced antenna systems which would include a much larger number of parameters per antenna and hence increase the number of agents in the graph. 

\clearpage

\printbibliography  

\appendix

The appendix contains additional details about the method, parameters used for the experiments and additional experiments.

\section*{A. Experimental details}

In this appendix, we describe the different hyperparameters for the algorithm and some implementation details about the experiments.
Our simulator models an LTE network with antennas that can be controlled remotely for changing the electrical downtilt angle and the maximum downlink transmitted power and operating at a frequency of \SI{2}{\giga\hertz}. 
The network has \num{10000} users uniformly distributed on the map, generating an average traffic of \SI{1}{Mbps} per cell. 

\begin{figure}
    \centering
    \includegraphics[width=\columnwidth]{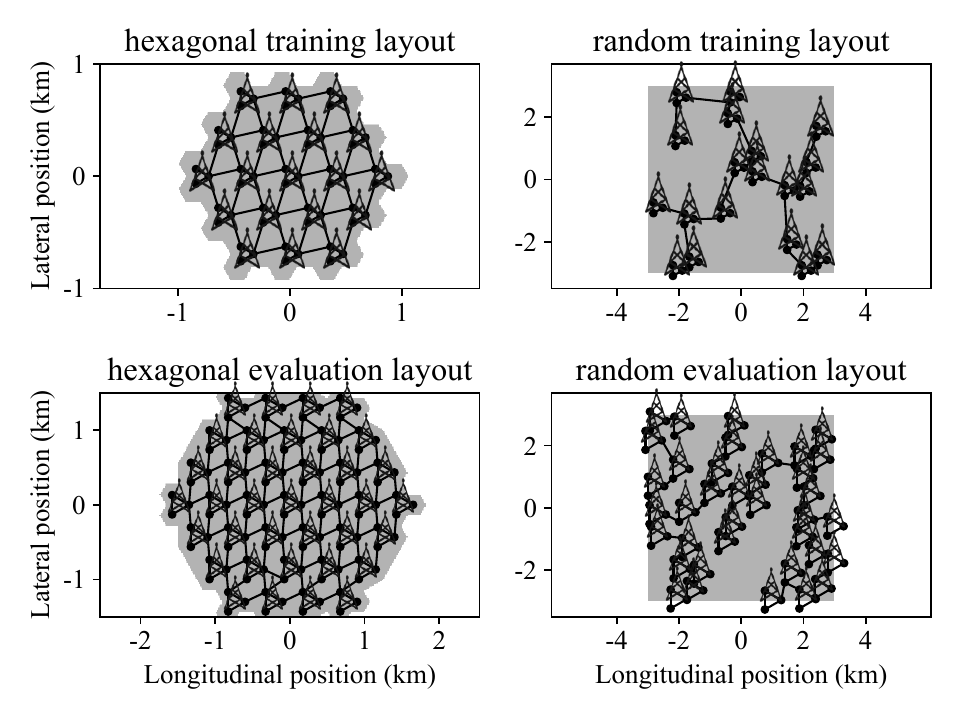}
    \caption{Visualization of the hexagonal and random deployments used for training and evaluation. Macro base station network deployments used in our experiments with the graph connecting the different cells (three per base station). The topology of the graph can change depending on the intersite distance. Evaluation layouts use denser deployments with more agents. The position of the node is offset from their real position for better visualization.}
    \label{fig:deployments-all}
\end{figure}

The training consists of \num{20000} steps split into episodes of \num{20} steps. 
At the beginning of an episode, a new deployment is sampled. We consider \num{19} base stations in the training environment, each consisting of \num{3} antennas, which makes a total of \num{57} cells. Each cell is associated to a learning agent. 
The average intersite distance between base stations is uniformly sampled between \SI{300}{\meter} and \SI{1500}{\meter}. 
We consider both hexagonal deployment and randomly generated deployment of the base stations with a minimum intersite distance, as illustrated in \cref{fig:deployments-all}. 
At the beginning of an episode, the electrical tilt of each antenna is reset to a random value within the range $[\SI{0}{\degree}, \SI{15}{\degree}]$. 
In the joint tilt and power control experiment the power is also reset to a random value while for the tilt experiment it is fixed at \SI{40}{\watt}.

The baselines are using exactly the same simulator configuration and are trained with the same random seed such that they experience exactly the same simulated networks both in training and evaluation.  
The observation features and the reward signal are normalized during training. 
They all use an $\epsilon$-greedy policy for exploration with a decay of the exploration rate, $\epsilon$, from \num{1} to \num{0.01} during the first half of the training.
Each training is repeated with three different random seeds.

\begin{table}[!t]
	\footnotesize
	\centering
	\caption{Hyperparameters for training the baselines and our algorithm. The first lines are common to all algorithms.}
	\label{tab:hyperparameters}
	\adjustbox{width=\columnwidth}{%
	\begin{tabular}{ll}
		\toprule
		$\gamma$ & \num{0.0} \\
		learning rate & \num{1e-3} \\
		initial $\epsilon$ & 1.0 \\
		final $\epsilon$ & 0.01 \\
		$\epsilon$ decrease steps & \num{10000} \\
		activation function & ReLU \\
		batch size & 64 \\
		\midrule
		(N)DQN architecture & FC(64), FC(32) \\
		\midrule
		GAQ architecture & GAT(32, 6 heads), GAT(32, 6 heads), FC(32), FC(32)  \\
		\midrule
		GQN architecture & GCN(32), GCN(32), FC(32), FC(32) \\
		GQN(GAT) architecture & GAT(32, 4 heads), GAT(32, 4 heads), FC(32), FC(32) \\
		GQN learning rate & \num{1e-2} \\
		\bottomrule
	\end{tabular}} 
\end{table}

All the baselines are implemented using rllib~\cite{liang2018} and the Pytorch geometric library~\cite{fey2019}, they all use a target network, prioritized replay, double Q-learning and distributed experience collection with \num{5} workers, each using \num{3} CPUs.
We report all the hyperparameters in \cref{tab:hyperparameters}. For all the baselines we carried out a hyperparameter search on the learning rate and number of layers and we rescaled the observation vectors and reward signals such that their value is between $[-1,1]$. 
We used a discount factor of \num{0.0} since our problem's goal, reaching the final optimal antenna configuration, does not require visiting intermediate states with performance degradation. 
Increasing the discount factor to \num{0.9} did not lead to any benefits for any of the baselines for this specific application. 
However, the method, as described in the previous section, still holds for larger discount factors.
It leads to the same asymptotic results but with slower convergence. 

In the N-DQN baseline, a maximum of \num{5} neighbors is considered. Each observation vector from the neighbors is stacked and fed to a feed forward neural network.

For our GQN algorithm, we tried adding encoding MLPs before the GNN layer but it did not bring any improvement in performance (nor did it damage it). 
We also experimented with a graph convolutional neural network layer instead of the graph attention one.

We provide a pseudocode of our proposed GQN method in \cref{alg:GQN}. It follows a similar off policy training procedure as DQN. We use a target network, double Q learning and prioritized experience replay. The state of each agent and the graphs are stored in the replay buffer.

\begin{algorithm}[h]
\begin{algorithmic}[1]
\State \textbf{Initialize}: GQN weights, replay buffer $\mathcal{D} = \{ \}$, training steps $T$, batch size $B$, $\epsilon$ schedule
\For{$t=1, \dots T$}
    \State Observe joint state $\mathbf{s}$, and adjacency matrix $A$
    \State With probability $\epsilon$, choose a random joint action $\mathbf{a}$
    \State Otherwise choose $\mathbf{a}=\mathbf{a^*}$ according to \cref{eq:igm}.
    \State Observe the next joint state $\mathbf{s}$, the next adjacency matrix $A'$, and the global reward $r$.
    \State Store the transition $(\mathbf{s}, A, \mathbf{a}, r, \mathbf{s'}, A')$ in $\mathcal{D}$.
    \State Sample batch $\{(\mathbf{s}^k, A^k, \mathbf{a}^k, r^k, \mathbf{s'}^k, A'^k) \, \text{for} \, k=1,\dots,B\}$ from $\mathcal{D}$.
    \State Assign $y^k = r^k + \gamma \max_{\mathbf{a'}}\sum_i Q_i(\mathbf{s'}^k, A'^k, a'_i) \forall k$ as per \cref{eq:Bellman}.
    \State Perform a gradient descent step on the loss ${1/B\cdot\sum_k[(\sum_i Q_i(\mathbf{s}^k, A^k, a^k_i) - y^k)^2]}$
\EndFor
\end{algorithmic}
\caption{GQN training procedure.}\label{alg:GQN}
\end{algorithm}

\section*{B. Additional Experiments}

\subsection*{Tilt Generalization}

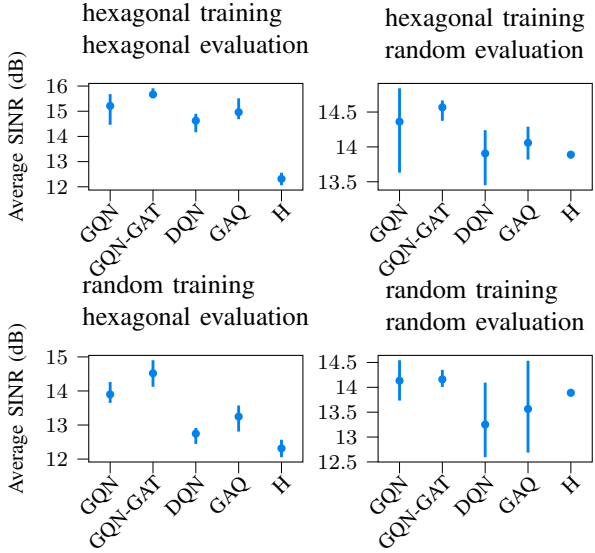
\begin{figure}[!t]
    \centering
\begin{tikzpicture}

\definecolor{darkgray176}{RGB}{176,176,176}
\definecolor{dodgerblue0130240}{RGB}{0,130,240}

\begin{groupplot}[group style={group size=2 by 2,vertical sep=2.2cm}, width=0.5\columnwidth, height=3cm]
\nextgroupplot[
tick align=outside,
tick pos=left,
title style={align=left},
title={hexagonal training \\ hexagonal evaluation},
unbounded coords=jump,
x grid style={darkgray176},
xmin=-0.5, xmax=4.5,
xtick style={color=black},
xtick={0,1,2,3,4,5},
xticklabels={GQN,GQN-GAT,DQN,GAQ,H},
x tick label style={rotate=45, anchor=north east, inner sep=0mm},
y grid style={darkgray176},
ylabel={Average SINR (dB)},
ymin=11.8712364860283, ymax=16.1040258501573,
ytick style={color=black}
]
table{%
x  y  draw  fill
0 15.2123838276273 0,130,240 0,130,240
1 15.6683761327389 0,130,240 0,130,240
2 14.6257836637029 0,130,240 0,130,240
3 14.9621204331694 0,130,240 0,130,240
4 12.314558025762 0,130,240 0,130,240
};
\addplot [only marks, mark size=1pt,mark=*, line width=1.08pt, dodgerblue0130240]
table {%
0 15.2123838276273
1 15.6683761327389
2 14.6257836637029
3 14.9621204331694
4 12.314558025762
};
\addplot [line width=1.08pt, dodgerblue0130240]
table {%
0 14.4633371385209
0 15.6787333835573
};
\addplot [line width=1.08pt, dodgerblue0130240]
table {%
1 15.5383183186699
1 15.911626333606
};
\addplot [line width=1.08pt, dodgerblue0130240]
table {%
2 14.1652532386594
2 14.8982876198943
};
\addplot [line width=1.08pt, dodgerblue0130240]
table {%
3 14.6838843464045
3 15.5116990049138
};
\addplot [line width=1.08pt, dodgerblue0130240]
table {%
4 12.0636360025796
4 12.5584725724441
};

\nextgroupplot[
tick align=outside,
tick pos=left,
title style={align=left},
title={hexagonal training \\ random evaluation},
unbounded coords=jump,
x grid style={darkgray176},
xmin=-0.5, xmax=4.5,
xtick style={color=black},
xtick={0,1,2,3,4,5},
xticklabels={GQN,GQN-GAT,DQN,GAQ,H},
x tick label style={rotate=45, anchor=north east, inner sep=0mm},
y grid style={darkgray176},
ymin=13.3812053999202, ymax=14.9102679285393,
ytick style={color=black}
]
table{%
x  y  draw  fill
0 14.3617878195756 0,130,240 0,130,240
1 14.5672959944874 0,130,240 0,130,240
2 13.9061944671223 0,130,240 0,130,240
3 14.0579773818793 0,130,240 0,130,240
4 13.8892051755826 0,130,240 0,130,240
};
\addplot [only marks, mark size=1pt,mark=*, line width=1.08pt, dodgerblue0130240]
table {%
0 14.3617878195756
1 14.5672959944874
2 13.9061944671223
3 14.0579773818793
4 13.8892051755826
};
\addplot [line width=1.08pt, dodgerblue0130240]
table {%
0 13.6304828132361
0 14.8407650863294
};
\addplot [line width=1.08pt, dodgerblue0130240]
table {%
1 14.3741246174767
1 14.6665155851318
};
\addplot [line width=1.08pt, dodgerblue0130240]
table {%
2 13.4507082421302
2 14.2400036315969
};
\addplot [line width=1.08pt, dodgerblue0130240]
table {%
3 13.8187777202919
3 14.2905632371882
};
\addplot [line width=1.08pt, dodgerblue0130240]
table {%
4 13.8348674721875
4 13.9176831068446
};

\nextgroupplot[
tick align=outside,
tick pos=left,
title style={align=left},
title={random training \\ hexagonal evaluation},
unbounded coords=jump,
x grid style={darkgray176},
xmin=-0.5, xmax=4.5,
xtick style={color=black},
xtick={0,1,2,3,4,5},
xticklabels={GQN,GQN-GAT,DQN,GAQ,H},
x tick label style={rotate=45, anchor=north east, inner sep=0mm},
y grid style={darkgray176},
ylabel={Average SINR (dB)},
ymin=11.9134688578056, ymax=15.044383670695,
ytick style={color=black}
]
table{%
x  y  draw  fill
0 13.8998679475545 0,130,240 0,130,240
1 14.5202266488269 0,130,240 0,130,240
2 12.7471026397811 0,130,240 0,130,240
3 13.246662925662 0,130,240 0,130,240
4 12.314558025762 0,130,240 0,130,240
};
\addplot [only marks, mark size=1pt,mark=*, line width=1.08pt, dodgerblue0130240]
table {%
0 13.8998679475545
1 14.5202266488269
2 12.7471026397811
3 13.246662925662
4 12.314558025762
};
\addplot [line width=1.08pt, dodgerblue0130240]
table {%
0 13.6514784774299
0 14.2640713587606
};
\addplot [line width=1.08pt, dodgerblue0130240]
table {%
1 14.1219356461126
1 14.9020693610182
};
\addplot [line width=1.08pt, dodgerblue0130240]
table {%
2 12.4445118843485
2 12.9143744870467
};
\addplot [line width=1.08pt, dodgerblue0130240]
table {%
3 12.8077595972763
3 13.5733008337825
};
\addplot [line width=1.08pt, dodgerblue0130240]
table {%
4 12.0557831674824
4 12.5667927648893
};

\nextgroupplot[
tick align=outside,
tick pos=left,
title style={align=left},
title={random training \\ random evaluation},
unbounded coords=jump,
x grid style={darkgray176},
xmin=-0.5, xmax=4.5,
xtick style={color=black},
xtick={0,1,2,3,4,5},
xticklabels={GQN,GQN-GAT,DQN,GAQ,H},
x tick label style={rotate=45, anchor=north east, inner sep=0mm},
y grid style={darkgray176},
ymin=12.4980469653835, ymax=14.6419089167212,
ytick style={color=black}
]
table{%
x  y  draw  fill
0 14.1339137367683 0,130,240 0,130,240
1 14.1588300117923 0,130,240 0,130,240
2 13.2511196236829 0,130,240 0,130,240
3 13.5645249401368 0,130,240 0,130,240
4 13.8892051755826 0,130,240 0,130,240
};
\addplot [only marks, mark size=1pt,mark=*, line width=1.08pt, dodgerblue0130240]
table {%
0 14.1339137367683
1 14.1588300117923
2 13.2511196236829
3 13.5645249401368
4 13.8892051755826
};
\addplot [line width=1.08pt, dodgerblue0130240]
table {%
0 13.7336685067439
0 14.5444606462059
};
\addplot [line width=1.08pt, dodgerblue0130240]
table {%
1 14.0072339985084
1 14.3512995616537
};
\addplot [line width=1.08pt, dodgerblue0130240]
table {%
2 12.5954952358988
2 14.0937314262046
};
\addplot [line width=1.08pt, dodgerblue0130240]
table {%
3 12.6867197183605
3 14.5345217022371
};
\addplot [line width=1.08pt, dodgerblue0130240]
table {%
4 13.8348674721875
4 13.9176831068446
};
\end{groupplot}

\end{tikzpicture}
    \caption{Generalization performance of the GQN algorithm and the baselines on deployments unseen at training time. The evaluation deployments are denser, as illustrated in \cref{fig:deployment}. The points represent the average performance and the error bars represent the 95th percentile confidence interval (sometimes smaller than the marker).}
    \label{fig:generalization-all}
\end{figure}

In the tilt generalization experiment, we tried different combination of training the models on random deployments and evaluating on hexagonal deployments and vice versa. The results are presented in \cref{fig:generalization-all}. In all cases, the GQN-GAT method present the best performance. 
An interesting outcome is that the GQN(GAT) model trained on random deployment, has a better performance on the hexagonal evaluation scenario (bottom left plot) than the DQN and GAQ models that were trained specifically on hexagonal topologies (top left plot). 
On the evaluation with random cell layout, the heuristic has a very close performance to some of the reinforcement learning algorithms, indicating that generalization to those scenarios is a difficult task, especially when the difference between training and deployment layout is large (top right plot). 

\subsection*{Joint Tilt and Power Control}

In this scenario the reward function is parameterized by $w$ which controls the trade-off between minimizing power and maximizing signal quality. We experimented with value of $w$ in $\{0.05, 0.1, 0.15, 0.2, 0.5\}$ for the reward functions. For the most extreme values for $w$ the agents learns to set the power to the maximum value (low $w$), or the minimum (high $w$). The values of $0.15$ and $0.2$ gave the most interesting results where the agent converged to non extreme power values. We witnessed the same effect for all methods and only report the results for $w=0.15$ for the sake of readability.

\begin{figure}[!t]
    \centering
    \input{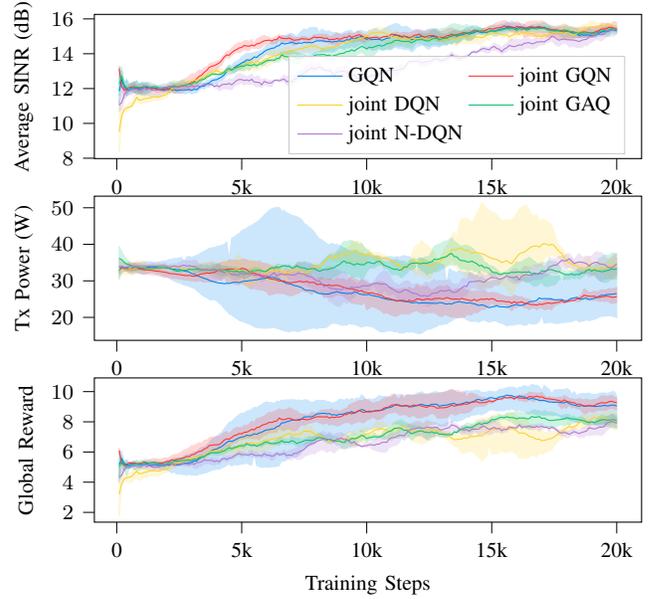}
    \caption{Training performance of GQN (including global reward) with separate agents for tilt and power, joint GQN and the baselines in the joint control environment. For a similar average SINR, GQN is able to yield a greater power reduction. The solid lines represent the mean over \num{3} random seeds and the shaded area is the \num{95}th percentile confidence interval.}
    \label{fig:power-tilt-training-all}
\end{figure}

In \cref{fig:power-tilt-training-all}, we add the evolution of the global reward. The GQN model are directly trying to optimize this reward while other algorithms must rely on local signals. As a consequence they do not give as much power reduction as GQN.

\section*{C. RSRP and SINR Calculations}

To derive the relation between SINR and cell configurations we first define the reference signal received power (RSRP) $R_{c,u}$ for a user $u$ connected to a cell $c$: 
\begin{equation}
R_{c_u}(\phi, \alpha) = \frac{P_c(\phi) G_{c,u}(\alpha)}{L_{c,u}}
\end{equation}
where $P_c(\phi)$ is the transmitted power of the antenna per reference signal resource element, as a function of the maximum transmitted power $\phi$. $G_{c,u}(\alpha)$ is the antenna gain which depends on the tilt angle $\alpha$ and the azimuth which is kept fixed in our model. $L_{c,u}$ is the path loss. 
To compute the antenna gains, one can use the relationship defined in the third generation partnership project~\cite{3gpp.36.814}. The maximum power $\phi$ is divided by the number of available resource blocks and multiplied by the cell specific reference signal power boost gain (set to one in our simulation).
The received power is only depending on the cell that the user is connected to and on the propagation environment which affects the path loss. 
The user also receives power from other cells which is regarded as interference in the SINR calculation.
For a user $u$, the cell $c$ that the user is attached to is assumed to be the cell yielding the largest received power.
The downlink SINR of a user $u$ is defined as the ratio between the received power from cell $c$ and the sum of the received power from all other cells, and the noise power $\mu$:
\begin{equation}
    \rho_{u}(\mathbf{\phi}, \mathbf{\alpha}) = \frac{R_{c_u}(\phi_c, \alpha_c)}{\sum_{i=1,i\neq c}^{N_{\text{cells}}} R_{i,u}(\phi_i, \alpha_i) + \mu}
    \label{eq:sinr}
\end{equation}
The noise power is calculated over a frequency bandwidth of one resource element (\SI{15}{\kilo\hertz}).
Improving the SINR of a user involves improving the received power as well as reducing interference from other cells.

\end{document}